\documentclass[aps,prd,twocolumn,superscriptaddress]{revtex4-1}

\usepackage{graphicx}
\usepackage{color}
\usepackage{longtable}
\usepackage{bm}
\usepackage{amsmath}
\begin{document}



\title{Direct dark matter search by annual modulation with 2.7 years of XMASS-I data}


\newcommand{\ICRR}{\affiliation{Kamioka Observatory, Institute for Cosmic Ray Research, the University of Tokyo, Higashi-Mozumi, Kamioka, Hida, Gifu, 506-1205, Japan}}
\newcommand{\IBS}{\affiliation{Center of Underground Physics, Institute for Basic Science, 70 Yuseong-daero 1689-gil, Yuseong-gu, Daejeon, 305-811, South Korea}}
\newcommand{\ISEE}{\affiliation{Institute for Space-Earth Environmental Research, Nagoya University, Nagoya, Aichi 464-8601, Japan}}
\newcommand{\IPMU}{\affiliation{Kavli Institute for the Physics and Mathematics of the Universe (WPI), the University of Tokyo, Kashiwa, Chiba, 277-8582, Japan}}
\newcommand{\KMI}{\affiliation{Kobayashi-Maskawa Institute for the Origin of Particles and the Universe, Nagoya University, Furo-cho, Chikusa-ku, Nagoya, Aichi, 464-8602, Japan}}
\newcommand{\Kobe}{\affiliation{Department of Physics, Kobe University, Kobe, Hyogo 657-8501, Japan}}
\newcommand{\KRISS}{\affiliation{Korea Research Institute of Standards and Science, Daejeon 305-340, South Korea}}
\newcommand{\Miyagi}{\affiliation{Department of Physics, Miyagi University of Education, Sendai, Miyagi 980-0845, Japan}}
\newcommand{\Tokai}{\affiliation{Department of Physics, Tokai University, Hiratsuka, Kanagawa 259-1292, Japan}}
 \newcommand{\Toku}{\affiliation{Insitiute of Socio-Arts and Sciences, The University of Tokushima, 1-1 Minamijosanjimacho Tokushima city, Tokushima, 770-8502, Japan}}
\newcommand{\YNU}{\affiliation{Department of Physics, Faculty of Engineering, Yokohama National University, Yokohama, Kanagawa 240-8501, Japan}}

\ICRR
\IBS
\ISEE
\Toku
\IPMU
\KMI
\Kobe
\KRISS
\Miyagi
\Tokai
\YNU

 \author{K.~Abe}\ICRR \IPMU    
  \author{K.~Hiraide}\ICRR \IPMU
  \author{K.~Ichimura}\ICRR \IPMU
  \author{Y.~Kishimoto}\ICRR \IPMU
  \author{K.~Kobayashi}\ICRR \IPMU
 \author{M.~Kobayashi}\ICRR 
  \author{S.~Moriyama}\ICRR \IPMU
  \author{M.~Nakahata}\ICRR \IPMU
  \author{T.~Norita}\ICRR \IPMU
  \author{H.~Ogawa}\ICRR \IPMU
  \author{K.~Sato}\ICRR 
  \author{H.~Sekiya}\ICRR \IPMU
  \author{O.~Takachio}\ICRR
\author{A.~Takeda}\ICRR \IPMU
  \author{S.~Tasaka}\ICRR
  \author{M.~Yamashita}\ICRR \IPMU
  \author{B.~S.~Yang}\ICRR \IPMU

 \author{N.~Y.~Kim}\IBS
  \author{Y.~D.~Kim}\IBS
  
 \author{Y.~Itow} \ISEE \KMI 
 \author{K.~Kanzawa} \ISEE
 \author{R.~Kegasa} \ISEE
 \author{K.~Masuda} \ISEE
  \author{H.~Takiya} \ISEE
\author{K.~Fushimi}
\altaffiliation{Present address: Department of Physics, Tokushima University, 2-1 Minami Josanjimacho Tokushima city, Tokushima, 770-8506, Japan}
\Toku
\author{G.~Kanzaki}
\Toku

\author{K.~Martens} \IPMU
\author{Y.~Suzuki} \IPMU
\author{B.~D.~Xu}\IPMU

  \author{R.~Fujita}\Kobe
  \author{K.~Hosokawa}
  \altaffiliation[Present Address:]{Research Center for Neutrino Science, Tohoku University, Sendai, Miyagi 980-8578, Japan}
  \Kobe 
  \author{K.~Miuchi}\Kobe
\author{N.~Oka}\Kobe
  \author{Y.~Takeuchi} \Kobe \IPMU
\author{Y.~H.~Kim} \KRISS \IBS
\author{K.~B.~Lee} \KRISS
\author{M.~K.~Lee} \KRISS

  \author{Y.~Fukuda} \Miyagi
   \author{M.~Miyasaka} \Tokai
  \author{K.~Nishijima} \Tokai
  \author{S.~Nakamura} \YNU

\collaboration{XMASS Collaboration}
\email{xmass.publications10@km.icrr.u-tokyo.ac.jp}
\noaffiliation

\date{\today}
\begin{abstract}
An annual modulation signal due to the Earth orbiting around the Sun would be one of the strongest indications of  the direct detection of dark matter.
 In 2016, we reported a search for dark matter by looking for this annual modulation  with our single-phase liquid xenon XMASS-I detector. That analysis resulted in a slightly negative modulation amplitude at low energy.
In this work, we included more than one year of  additional  data,  which more than doubles the  exposure  to 800 live days with the same 832 kg target mass.
When we assume weakly interacting massive particle (WIMP) dark matter elastically scattering on the xenon target, 
the exclusion upper limit for the WIMP-nucleon cross section  was improved by a factor of 2 to 1.9$\times$10$^{-41}$cm$^2$ at 8 GeV/c$^2$ at 90\% confidence level with our newly implemented data selection through a likelihood method.
  For the model-independent case, without assuming any specific dark matter model, we obtained more consistency with the null hypothesis  than before with a $p$-value of 0.11 in the 1$-$20 keV energy region. This search probed this region with an exposure that was larger than that of DAMA/LIBRA.
We also did not find any significant amplitude in the data for periodicity with periods between 50 and 600 days in the energy region between 1 to 6 keV.
\end{abstract}

\pacs{95.35.+d,29.40.Mc, 14.80.-j}

\maketitle


\section{Introduction}

Although we do not yet know what dark matter is, its existence is well established. 
Various approaches are used to uncover its nature in direct and indirect searches as well as in collider experiments \cite{PDG}.
 The Earth's velocity relative to the dark matter distribution in the Galaxy  changes as the Earth moves around the Sun and would thus produce modulation with a maximum in June at the level of a few percent in a putative dark matter signal rate if it were  observed with terrestrial detectors \cite{Drukier}. 
  The DAMA/LIBRA experiment observed annual modulation of its event rate with a 9.3$\sigma$ significance in 1.33 ton$\cdot$year of data taken over 14 annual cycles with 100 to 250 kg of NaI(Tl)  \cite{dama}.  
 An interpretation of the result as a dark matter signature has been in question  for more than a decade because of tension with  experiments using  other target materials.

 Weakly interacting massive particles (WIMPs) are still well motivated  among the many candidates for dark matter particles to date, however,  the WIMP hypothesis appears inconsistent with results from experiments that report signals interpreted as WIMP dark matter \cite{KOPP_DAMA}.
 In particular two-phase liquid xenon time projection chambers (TPC) such as  XENON1T \cite{XENON1T}, LUX \cite{LUX}, PandaX-II \cite{PANDA}, and ZEPLIN-III \cite{ZEPIII} have consistently published null results for nuclear recoil based WIMP searches that would exclude the DAMA modulation finding if it were of that origin. Interpreting DAMA as dark matter-electron scattering and searching for electron recoil based modulation in other dark matter detectors has thus become more interesting as they can produce keV energy deposition in the detector as observed by  DAMA/LIBRA while avoiding other direct detection constraints \cite{KOPP, FELDSTEIN,FOOT}.
 
  XMASS-I, a single-phase liquid xenon (LXe) detector, has a high light yield and low background. XMASS probed this possibility and looked for signal not only from nuclear recoils but also from electrons and gamma rays emanating from interactions of other dark matter candidates such as axionlike particles and super-WIMPs  as well as solar Kaluza-Klein axions \cite{XMASS1, XMASS_SW, XMASS4}. 
   In 2016, XMASS published an annual modulation search for dark matter and a small negative amplitude  was found in the 359.2 live days of data between November 2013 and March 2015 with $p$-values of 0.014 or 0.068 for the two different analyses reported in \cite{XMASS_MOD}.  Since then we have taken more than another year of data with more stable detector conditions in terms of temperature, pressure, and scintillation light yield resulting in a total live time of 800.0 days.
XMASS has only a modest background comparable to that of  DAMA/LIBRA; it has a large target  mass of 832 kg  LXe and the  total exposure  of 1.82 ton$\cdot$ year is larger than that of the DAMA/LIBRA experiment.
Recent annual modulation searches were reported by XENON100 \cite{XENON_MOD} without discriminating against electron events and by DM-Ice, also with a NaI(Tl) target \cite{DM-ICE}.  These detectors are located at the Gran Sasso laboratory in Italy and the South Pole, respectively.
Compared to these other experiments,  XMASS  has the lowest energy threshold (1 keV) and also looks for modulation in both a different geographical location as well as at a different underground site.

\section{The XMASS experiment}
XMASS-I employs a single-phase LXe detector that observes only the scintillation light from LXe  and has no  electric field.
 The detector is located at  the Kamioka Observatory in Japan, which is  an underground laboratory with  an overburden of 1,000 m rock (2,700-meter water equivalent). The detailed design and performance  of the detector are described in \cite{XMASS_Det}. The detector is immersed in a water tank, 10 m in diameter and 10.5 m in height, which is equipped with 72 Hamamatsu R3600 photomultiplier tubes (PMTs) and acts as an active muon veto and a passive radiation shield against neutrons and gamma rays from the surrounding rock.  
A vacuum insulated inner copper vessel holds about 1.1 ton of  LXe and  642 high quantum efficiency (28\%$-$40\% at 175 nm) Hamamatsu R10789 PMTs are mounted on the inner surface of the LXe detector, which has a pentakis-dodecahedral shape that approximates a sphere with an average radius of 40 cm and contains 832 kg of the LXe. 
 The number of nonoperational PMTs during the relevant data taking period rose from 7 to 9  as two more PMTs developed  high rate dark noise or electrical problems.

\begin{figure}[t]
\centering
\includegraphics[width=0.52\textwidth]{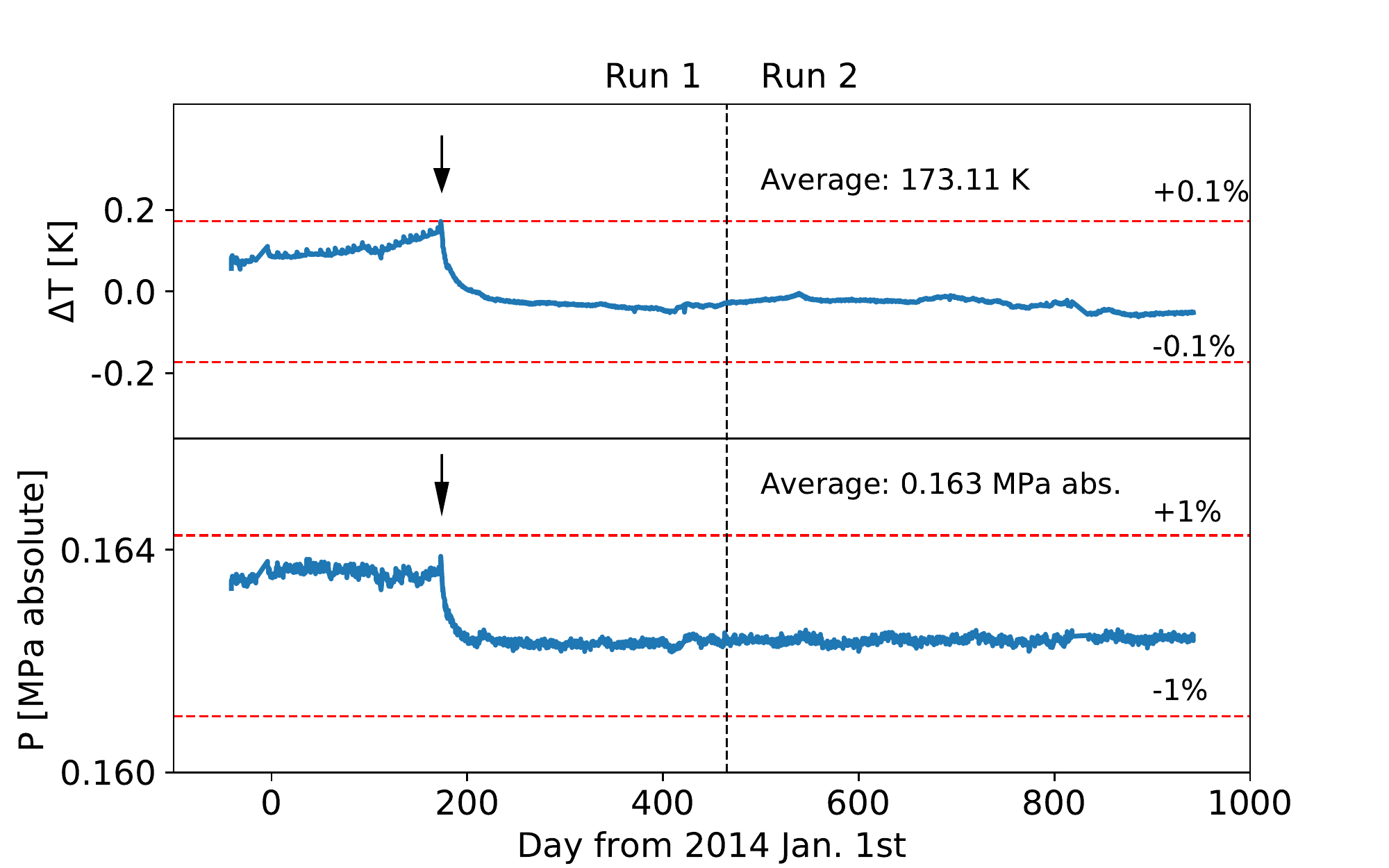}
\caption{Time evolution of the temperature deviation from the average ($\Delta$T) and pressure ($P$) of the detector  over time. The temperature drop at the 174th day (arrow) was due to a change of the reference temperature sensor for the refrigerator.}
\label{fig:p-t}
\end{figure}

   Starting from the previous annual modulation data set between November 2013 and March 2015 \cite{XMASS_MOD},  we have added data  taken between April 2015 and July 2016. Hereafter, we call the former period  run~1 and the later period run~2. 
   Figure \ref{fig:p-t} shows the stability of the detector  temperature measured in the LXe and the pressure above the liquid over that whole time. The average temperature and absolute pressure  were 173.11 K and 0.163 MPa, respectively.  The temperature drop of about 0.2 K  at the 174th day was due to a change of the reference temperature sensor for the feedback loop that controls the detector temperature. However, this temperature change causes only a negligible change in LXe density and  had no  impact on the  LXe scintillation light yield as shown in the top panel of Fig.\ref{fig:stability}.
  Note that we recovered some  small data set  within the run 1 period and added about 28.6 days to run 1 over the data set of the previous paper \cite{XMASS_MOD}. The live times of run 1 and run 2 are 387.8 and 412.2 days, respectively. The total live time thus became 800.0 days, with the total exposure becoming 1.82 ton$\cdot$year as summarized in Table~\ref{tab:sum}.
  
 \begin{table*}
   \caption{Summary of XMASS-I data exposures and PE yield stability obtained from  the regular $^{57}$Co calibrations. }
  \begin{tabular*}{1.0\textwidth}{@{\extracolsep{\fill}} llllll @{}}
    \hline
      \hline
    &Date &  Live time [day]  &  Exposure [ton$\cdot$year]  & PE yield stability\\ 
    \hline
    Run 1& Nov/20/2013 - Mar/31/2015 &387.8 & 0.884& $\pm$2.4\% \\ 
    Run 2& Apr/1/2015 - Jul/20/2016 &412.2 & 0.940&  $\pm$0.5\% \\ 
    \hline
      Total&& 800.0 &  1.82\\ 
    \hline
          \hline
 \end{tabular*}

  \label{tab:sum}
  \end{table*}
\section{Calibration}
\subsection{PMT gain and energy calibration}
 We used the same calibration procedures as in \cite{XMASS_MOD}.  
The  PMT gain  was monitored by means of the single photoelectron (PE) signal from a low-intensity blue LED embedded  in the inner detector wall. The scintillation light yield  was tracked by inserting a $^{57}$Co source  into the detector  every one or two weeks \cite{XMASS_Det, XMASS_Cal}.  The $^{57}$Co calibration data (122 keV gamma rays) were taken  at 10 cm intervals from $z=-$40 cm to $+$40 cm (9 locations in total) along the central vertical axis of the detector to track the PE yield and optical properties of the LXe. The number of events for each source position was about 20,000. The position dependence of the PE response was  about 10\% along the vertical axis from the detector wall to the detector center and this was well reproduced by the XMASS detector Monte Carlo (MC) simulation within $\pm$3\%. 

\subsection{PE yield stability}
\begin{figure}[b]
\centering
\includegraphics[width=0.5\textwidth]{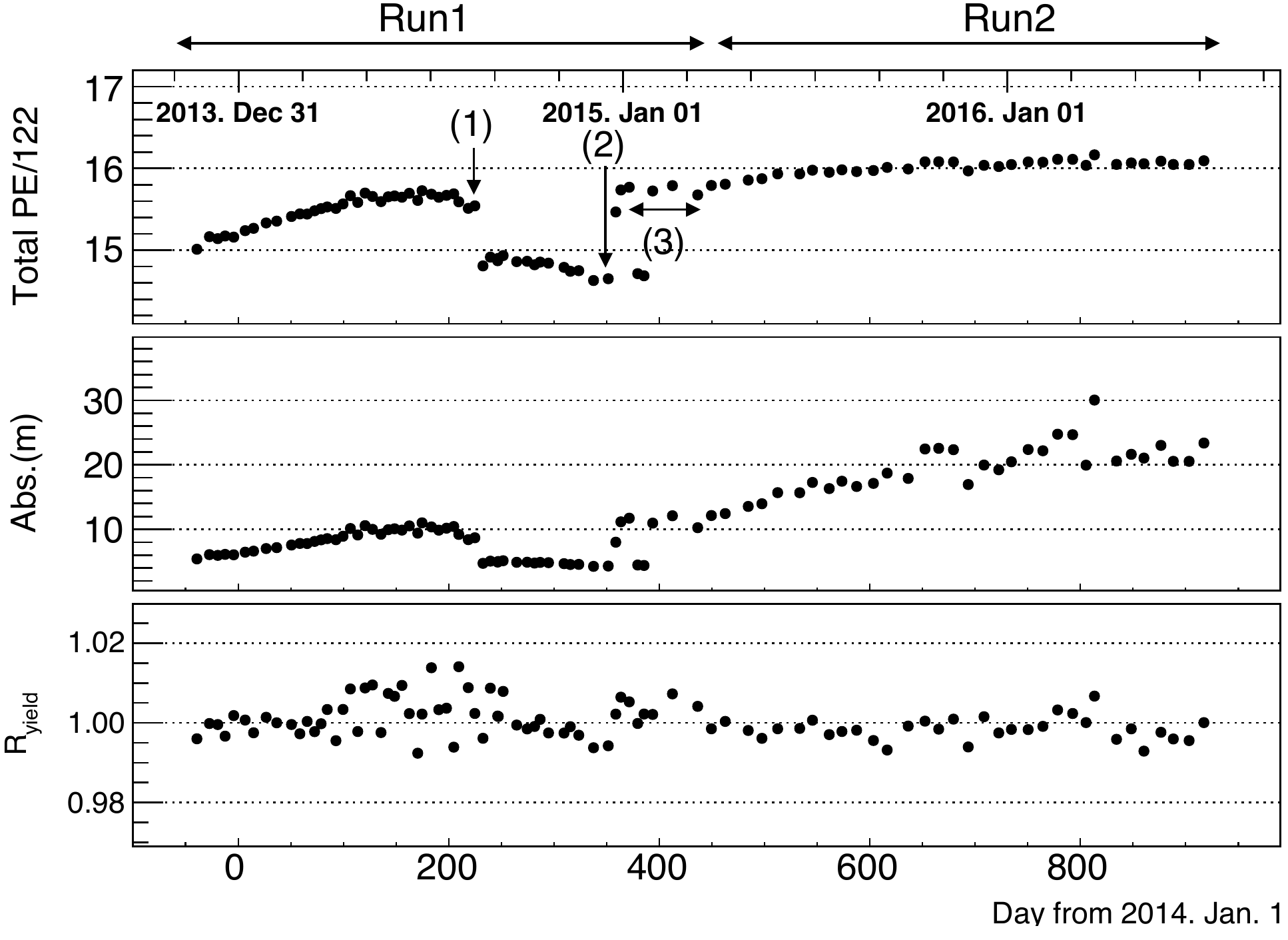}
\caption{ The PE yield  was  monitored with 122 keV gamma rays from  a $^{57}$Co source (top panel). 
 Events that led to abrupt changes: (1) A power failure on 17, August 2014. (2) Toggled to the other cold finger at the detector.
(3) Warming up those cold fingers while extracting the gas surrounding them. After that, we started to circulate the gas through the API getter.
  The absorption length for the scintillation light  and the relative intrinsic scintillation light yield ($R_{yield}$) were evaluated with the help of the XMASS MC  from $^{57}$Co calibration data.}
\label{fig:stability}
\end{figure}

 The PE yield at 122 keV (total PE/122) was monitored with the $^{57}$Co calibration and  appropriately weighted over the entire volume. It is shown in the top panel of Fig.~\ref{fig:stability}.  This time the dependence of the PE yield was taken into account in our analysis by linearity interpolating between calibrations.
 The absorption and scattering length for scintillation light as well as the intrinsic light yield of the liquid xenon scintillator are inferred from the $^{57}$Co calibration 
 data at 9 different positions from $z$ = $-$40 cm to $z$ =$+$40 cm by matching the PE hit patterns in data and MC \cite{XMASS_Det}. The scattering length remained stable at around 52 cm.  The time variation of the absorption length and the intrinsic light yield  are shown in the lower two panels of Fig.~\ref{fig:stability}. 
 
  The standard deviation of the PE yield  during run 1 was $\pm$ 2.4\%.  It changed gradually from the beginning of run 1, however, with the following features standing out: 
  (1)~It suddenly dropped  after a power failure on  August 17, 2014, during which the detector was cooled by liquid nitrogen through a cooling coil attached to the inner vessel. (2)~Later, sharp PE yield changes were seen again when we toggled between cold fingers as the operation was swapped from one of the two pulse tube refrigerators to the other for maintenance in December 2014.
  (3)~Finally, after warming up both cold fingers to room temperature while extracting the gas surrounding the cold fingers, the previous best PE yield was recovered and good stability was achieved after starting gas circulation through an API hot zirconium getter (NIPPON API Co., Ltd.) in March 2015.
  According to XMASS MC studies with $^{57}$Co calibration data, those changes can be explained by changes in the scintillation light absorption length in LXe. 
  To explain the data this absorption length has to vary from about 4 m to 30 m. We think that  impurities such as water, nitrogen, and oxygen caused the observed total PE changes.   The standard deviation of the PE yield was  only $\pm$0.5\% in run 2.
 The relative intrinsic light yield ($R_{yield}$) of  the LXe scintillator stayed within $\pm 0.6\%$ and $\pm$0.3\%  in run 1 and Run 2, respectively.
 
\subsection{Energy scale}
 In this paper, we use two different energy scales: (1) keV$_{\rm ee}$ represents an electron equivalent energy incorporating all the gamma-ray calibrations in the energy range between 5.9 and 122 keV. For these calibrations,  $^{55}$Fe, $^{109}$Cd,  $^{241}$Am, and $^{57}$Co  sources were  inserted  into the sensitive volume of the detector.  The nonlinearity of  the energy scale was taken into account in those calibrations using the model from Doke {\it et al}. \cite{DOKE}.  Recently, the energy scale  below 5.9 keV was confirmed with the escape X-ray peak in the $^{55}$Fe calibration which has a weighted mean energy of 1.65 keV. The scintillation efficiency at this energy was about 40\% smaller than that at 122 keV, with an uncertainty of 10\% and the energy scale in this analysis is within this error.
 (2) keV$_{\rm nr}$ denotes the nuclear recoil energy which is estimated from the observed PE count using a non-linear response function anchored at 122 keV for zero electric field from \cite{Leff}. The energy threshold for the analysis in this paper  corresponds to 1.0 keV$_{\rm ee}$ or 4.8 keV$_{\rm nr}$.

\section{Data Analysis}
\subsection{Data selection}
\begin{figure*}[htbp]
\centering
\includegraphics[width=1.0\textwidth]{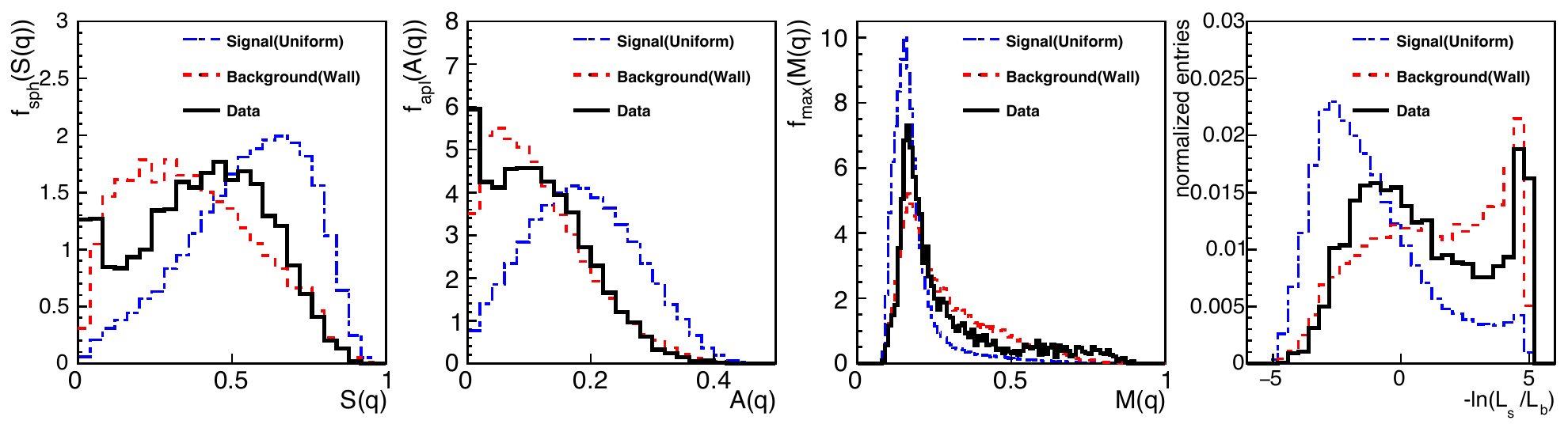}
\caption{The probability density functions $f_{sph}( S(\bm{q})), f_{apl}(A(\bm q))$ and  $f_{max}(M(\bm q))$} at 2.0$\pm$0.1 keV$_{\rm ee}$ (left three figures) evaluated at the 6 m absorption length as we had at the beginning of run 1.  Uniformly distributed events in the detector (blue) and events within about 3 cm from the detector wall (red) are shown together with data (black).  The log likelihood ratios --ln$L_{s}/L_b$  are also shown to the right.
\label{fig:like}
\end{figure*}

 Before retrieving time variation information from the data, event reduction was performed  to reduce background mainly from Cherenkov light in PMT windows and from events near the detector wall as described in \cite{XMASS_MOD} by applying standard cuts. Events with 4 or more PMT hits in a 200~ns coincidence timing window without a muon veto tagged as ``ID Trigger" by the data acquisition system were initially selected.
  A ``Timing cut" was applied that rejects events occurring within 10 ms from the previous event  and having a standard deviation in their hit timings of greater than 100~ns. This cut avoids events caused by afterpulses of bright events induced by, for example, high energy gamma rays or alpha particles.  A ``Cherenkov cut''  removed events which produce light predominantly from Cherenkov emission, in particular from the beta decays of $^{40}$K in the PMT photocathode. Events for which more than 60\% of their PMT hits arrive in the first 20 ns were classified as Cherenkov-like events \cite{XMASS1}.
  Finally, we construct a likelihood function ($L$) to remove background events that occurred in front of a PMT window or near the detector wall based on PE hit patterns in the event.
 The sphericity and aplanarity of  events have been used in high energy physics to find jets, for instance in \cite{HANSON};  we calculated these parameters based on the observed PE distribution in an event: 
\begin{equation}
 \label{eq:like}
  L = f_{sph}( S(\bm{q}))  \times f_{apl}(A(\bm q)) \times f_{max}(M(\bm q))
\end{equation}

where \(\bm q=(q_1, ..., q_{642})\) is the number of PE for all 642 PMTs in one event.  The number of PE for nonoperational PMTs was set to zero. \(S(\bm q)\), \(A(\bm q)\), \(M(\bm q)\) are the parameters of sphericity, aplanarity, and the maximum in $\bm q$  for the event over the sum of the $q_\alpha$ where $\alpha$  runs overall  PMTs, respectively.
\(f_{sph}\),   \(f_{apl}\), and  \(f_{max}\) are probability density functions for those parameters and  will be described in more detail later.

The sphericity tensor \(S^{ij}\) of an event is defined as
\begin{equation}
 \label{eq:1}
S^{ij} = \frac{\sum\limits_\alpha q_{\alpha}^i q_{\alpha}^j}{\sum\limits_\alpha q^2_\alpha}, 
\end{equation}
where $i,j$ = 1, 2, 3 correspond to the $x, y,$ and $z$ components by taking the detector center as the origin.  \(q_{\alpha}^{i}\) is  the $i$th component of the PE weighted  vector pointing from the detector center to \(\alpha\)th PMT. \(S^{ij}\) has three eigenvalues \(\lambda_1 \ge \lambda_2 \ge \lambda_3\)  (\(\lambda_1 + \lambda_2 + \lambda_3 = 1\)) and the sphericity \(S\) of the event
is defined as:
 \[ S(\bm q) = \frac{3}{2}(\lambda_2 + \lambda_3). \]

If the event topology is perfectly spherical,   $S(\bm q)$ becomes 1 and if  the event topology degenerates into a line, $S(\bm q)$ becomes 0.
 
The aplanarity \(A\) is defined as

\[ A(\bm q) = \frac{3}{2} \lambda_3.\]

Therefore, if the event is contained in a plane, \(A(\bm q) = 0\).  For a perfect sphere, \(A(\bm q) = \frac{1}{2}\).

The maximum PE fraction \(M\) for an event is defined as
\[M(\bm q) = \frac{q_{max}}{\sum\limits_\alpha q_{\alpha}},\]
where \(q_{max}\) is the maximum of the PE values of the PMTs in that event.

\begin{figure}[b]
\centering
\includegraphics[width=0.5\textwidth]{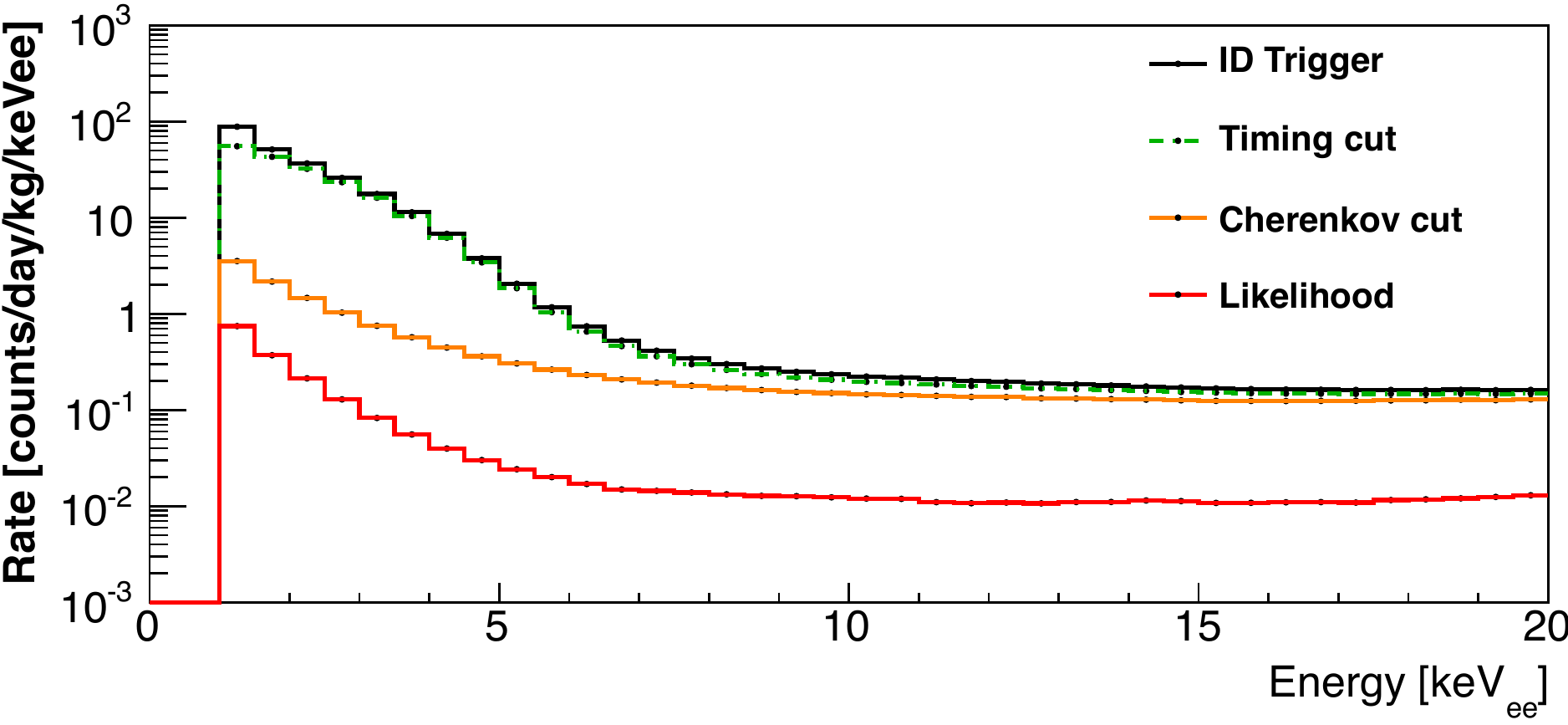}
\includegraphics[width=0.5\textwidth]{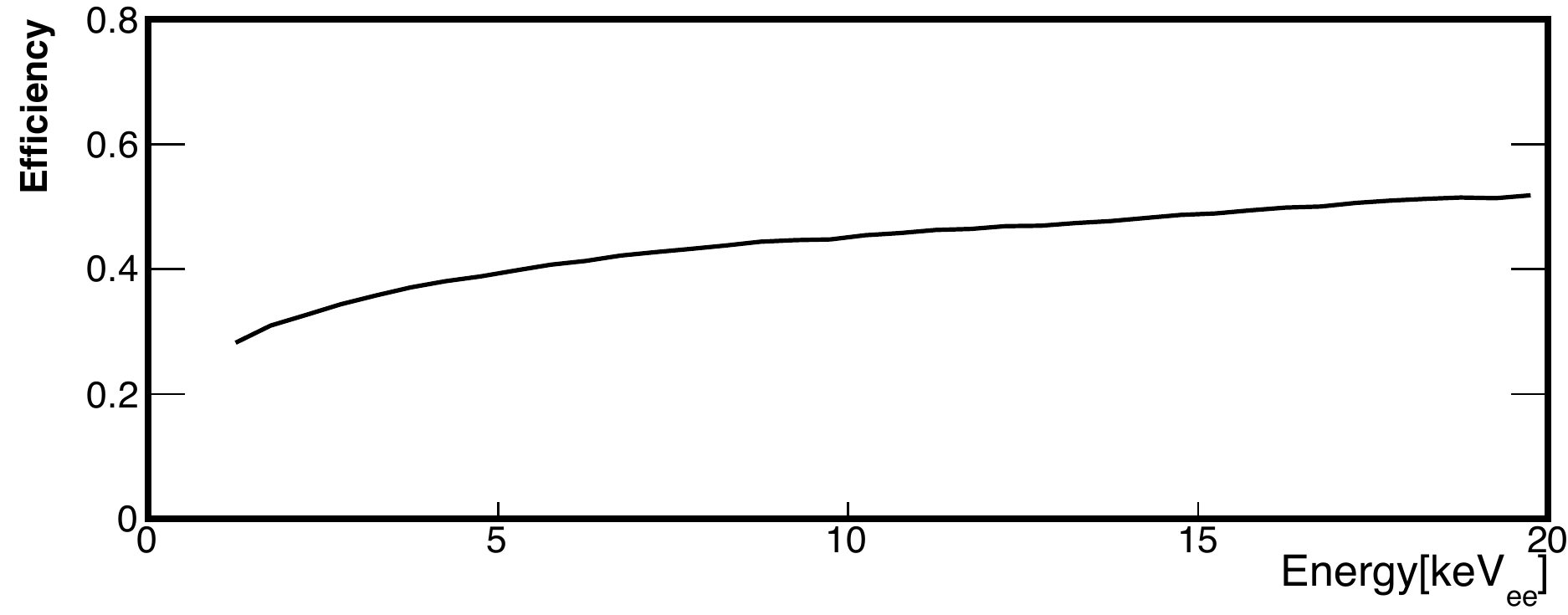}
\caption{Energy spectrum after the event selection steps for the whole data sample before efficiencies correction (top). Overall efficiency for uniformly distributed signal events after applying all cuts including our likelihood with 832 kg LXe mass (bottom).
The energy threshold in this analysis is 1.0 keV$_{\rm ee}$.}
\label{fig:hist}
\end{figure}

  To construct a likelihood ratio that allows discriminating against events entering the final sample from the vicinity of the wall, we created two samples of MC events. One sample was called the signal-like sample. It is uniformly distributed  throughout the detector volume and is used to obtain the likelihood function $L_s$  following  Eq.~(\ref{eq:like}).
The other sample used for  $L_b$ contains events from an otherwise uniform MC sample that were closer than 3 cm to the wall; these we considered backgroundlike events. 
 We obtained  the probability density functions for the $S(q)$, $A(q)$, and $M(q)$ required for Eq.~(\ref{eq:like}) after event reduction by the standard cuts was applied to each sample.
   Figure \ref{fig:like} shows  probability density functions for these three variables  and the resulting likelihood ratio for the energy  2.0$\pm$0.1~keV$_{\rm ee}$, together with the same functions for the data in the case of 6 m absorption length at the beginning of run 1. 
   The total observed PE response of the PMTs inside the detector  is understood at the level of $\pm$3\% between the data and MC for our $z$-dependent $^{57}$Co calibrations.
This choice of background sample was made in light of the complexity of modeling background in the immediate vicinity of the ID inner surface \cite{XMASS_FV}.  Thus we simply used the above background-like sample and considered the impact of the simplification by considering appropriate systematic errors. The cut parameter in $-$ln($L_{s}/L_{b}$) was chosen to keep 50\% efficiency after the standard cuts. Its distributions  for the energy 2.0$\pm$0.1~keV$_{\rm ee}$  are shown in the last panel of  Fig.~\ref{fig:like}. 
 To maintain the signal efficiency, the cut value of $-$ln($L_{s}/L_{b}$) is dependent on the  observed number of PE. At the 2.0$\pm$0.1~keV$_{\rm ee}$, for example,  events with   $-$ln($L_{s}/L_{b}$) $\leq$  -1.6 were kept  for further analysis.
 
The top panel of Fig.~\ref{fig:hist} shows the energy spectrum after each event selection step for  the whole data sample (run 1 and run 2) before efficiency correction.
 The count rate for data after all cuts including our likelihood cut is $\sim$0.75 events/day/kg/keV$_{\rm ee}$ at 1.0  keV$_{\rm ee}$.  
  The signal efficiency was evaluated from MC simulation with events uniformly distributed throughout the sensitive volume.  In order to estimate the efficiency, a flat energy spectrum was assumed and the fraction of remaining events after all cuts was calculated.  The bottom panel of  Fig.~\ref{fig:hist} shows the signal efficiency after all cuts with 832 kg LXe target. Overall our improved event selection $-$ while keeping the signal efficiency$-$ brings about a further reduction in data size by about 30\% at low energy compared to our previous publication.

\begin{figure*}[htbp]
\centering
\includegraphics[width=1.0\textwidth]{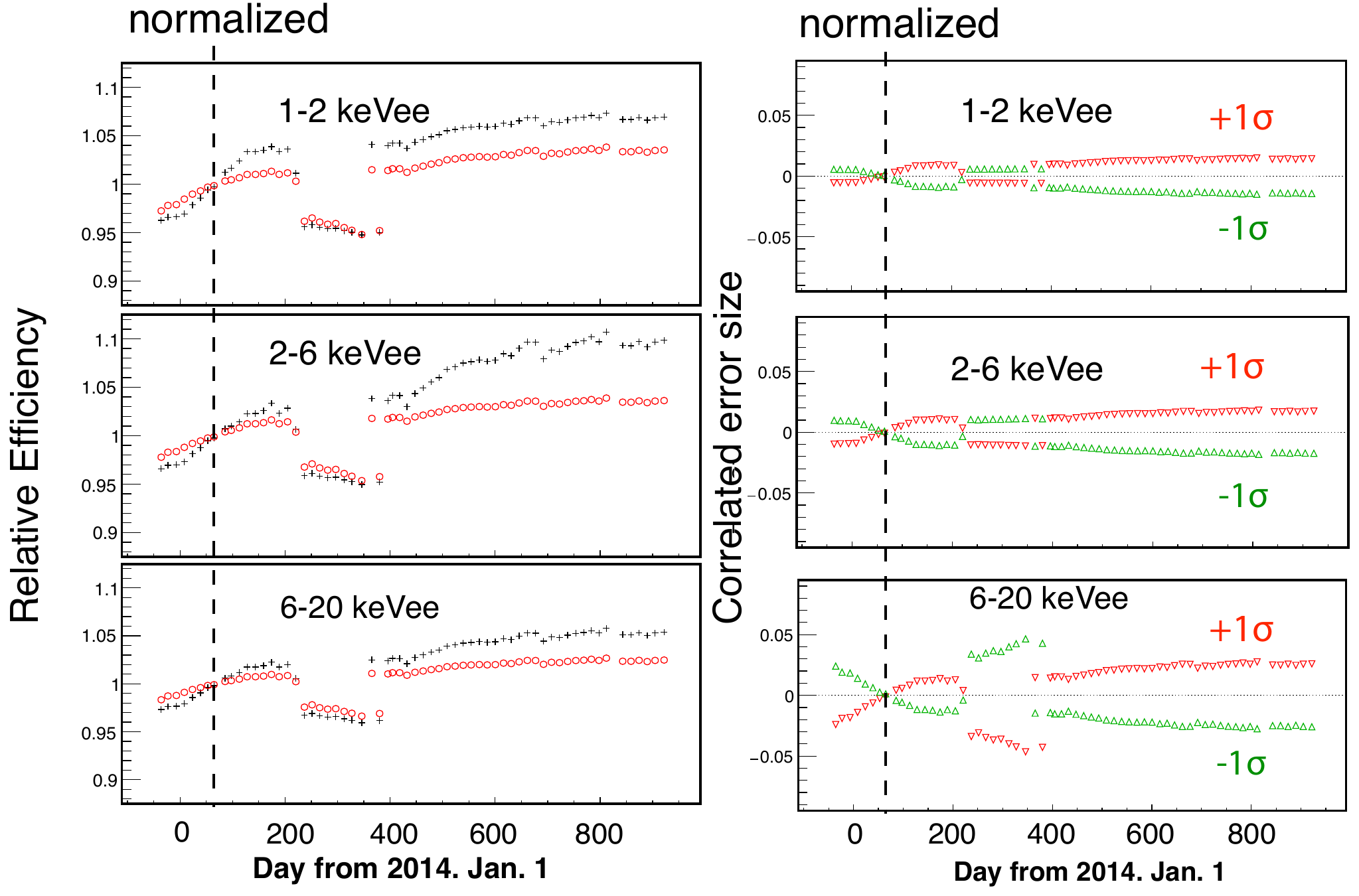}
\caption{ Relative efficiencies mean value for both signal (circle) and background (cross) events: We normalized  the overall efficiency at an absorption length of 8 m for different energy ranges (left).  1$\sigma$ ranges for the error of the relative efficiencies shown on the left are shown on the right.}
\label{fig:sys}
\end{figure*}

\subsection{Systematic errors}
\label{sec:sys}
 Systematic errors associated with PE yield changes during exposure were treated in the same way as described in \cite{XMASS_MOD}. 
 We found that the primary radioactive background in the low energy region came from the aluminum seal of our  PMT windows and secondary  gamma rays from the PMT body \cite{XMASS_Det}. The radioactivity in the  PMT aluminum seal was measured by the study with an HPGe detector and the main components are  $^{238}$U and  $^{210}$Pb with 1.5$\pm$0.4 and 2.9$\pm$1.2 Bq  for all PMTs in the detector, respectively. The scintillation light from their $\beta$ rays is emitted right at the PMT seals and its absorption on the way to the opposite side of the inner detector is the key parameter affecting the background level per energy bin.
We evaluated the absorption length dependence of the relative cut efficiencies based on this background.  MC simulation of events from the PMT aluminum and body radioactivity was used to estimate the relative efficiencies and their uncertainties used in our systematic errors.  More details of our  background study, including uncertainty in aspects of aluminum seal geometry were given in \cite{XMASS_FV}. 
As for internal background from radioactive isotopes in LXe, we identified several isotopes in studies towards our two-neutrino double electron capture search paper \cite{XMASS_DEC}. Those were  $^{222}$Rn daughters (8 mBq), $^{85}$Kr (0.2~mBq), $^{39}$Ar (0.6 mBq), $^{14}$C (0.2 mBq) per 832 kg LXe.  These rates  changed after changes in the gas circulation from March 2015 and this background survived data reduction because it was uniformly distributed in the detector.  However, the total  rate was of an order of 10$^{-4}$ events/day/kg/keV$_{\rm ee}$ and  it turned out to be a negligible contribution to the overall background.
	
　To treat the energy dependence of the relative efficiencies for both signal and background events, the energy range 1$-$20 keV$_{\rm ee}$ was divided into 3 energy bins: 1$-$2 keV$_{\rm ee}$,  2$-$6 keV$_{\rm ee}$, and 6$-$20 keV$_{\rm ee}$. The mean of relative efficiencies and their error size as a function of time  from January 2014 are  shown Fig.~\ref{fig:sys} (left) and (right), respectively. The mean relative efficiency in the 1--20 keV$_{\rm ee}$ energy range vary from  $-$5\% to +10\% for the background events and from about $-$5\% to +4\% for the signal events over the relevant absorption length range.  These efficiencies vary in the range from 0.01 to 0.05 relative to each other for all energy and time ranges. As we normalized the relative efficiency and the size of these errors at an absorption length of 8 m in this analysis, the relative efficiency, and the correlated error became one and zero at 70 days,  respectively.
  Note that these errors affect the count rate of the final data samples and are correlated between energy bins as well as time period bins because the PE yield, and with it the energy scale,  depends on time.
  This relative efficiency is the dominant systematic uncertainty in the present analysis. The second largest contribution comes from a gain instability of the waveform digitizers (CAEN V1751)  between April 2014 and September 2014 introduced by  a different calibration method for the digitizers that was used only  during that period. This latter systematic contributes an extra uncertainty of 0.3\% to the energy scale. Other contributions from LED gain calibration, trigger threshold stability, and timing calibration were negligible. 
  
\section{Results and Discussion}
 To obtain the annual modulation amplitude from the data, a least squares method for time-binned data was used to fit both run~1 and run~2 simultaneously. The data set was divided into 63 time-bins ($t_{bins}$) with roughly 15 days of real time each. The data in each time-bin was  further divided into energy-bins  ($E_{bins}$) with a width of 0.5 keV$_{\rm ee}$.  A pull  method \cite{PULL} was used to fit all energy- and time-bins simultaneously and  treat the correlated errors above.  We performed two analyses, one assuming WIMP interactions and the other independent of any specific dark matter model. Hereafter we call the former case the WIMP analysis and the latter the model-independent analysis. 
  \subsection{WIMP analysis} 
 In the case of the spin-independent WIMP analysis,  $\chi^{2}$ is defined as:
\begin{equation}
\chi^2 = \sum\limits_{i}\limits^{E_{bins}} \sum\limits_{j}\limits^{t_{bins}} 
\left(\frac{(R^{{\rm data}}_{i,j}-R^{\rm ex}_{i,j}(\alpha,\beta))^2}{\sigma({\rm stat})^2_{i,j}+\sigma({\rm sys})^2_{i,j}}\right)+\alpha^{2}+\sum^{Nsys}\beta_i^{2}, 
\end{equation}

where $R_{i,j}^{\rm data}$, $R_{i,j}^{\rm ex}$,  $\sigma(\rm{stat})$$_{i,j}$, and $\sigma(\rm{sys)}$$_{i,j}$ are the data rate and expected MC event rate and the  statistical and the systematic errors of the expected event rate for  the $i$th energy and $j$th time bin, respectively. Time is denoted as the number of days from January 2014. The penalty term $\alpha$ relates to the  overall size of the relative efficiency error and it is common for all energy bins; therefore the size of  their error simultaneously scales  with $\alpha$  in the fit procedure.  $\alpha$=1($-$1) corresponds to the $1\sigma(-1\sigma)$ correlated systematic error as shown in Fig.~\ref{fig:sys} (right) on the expected event rate, $R^{\rm ex}_{i,j}(\alpha,\beta)$, in that energy bin.  $\alpha$ is determined during the minimization of $\chi^2$  and increases $\chi^2$ by  $\alpha^2$.
The other penalty term, $\beta_i$, relates to the systematic uncertainty of the expected WIMP signal simulation. This uncertainty has two main components: the scintillation efficiency \cite{Leff} and the time constant of nuclear recoil signals. A time constant of 26.9~$^{+0.8}_{-1.2}$ nsec was used based on a neutron calibration of the  XMASS-I detector \cite{XMASS_NEUTRON}.
The expected signals are simulated with parameters corresponding to the limits of the 1$\sigma$ error range to estimate the impacts on the amplitude $A^{s}_{i}(\beta)$ and the unmodulated component $C^{s}_{i}(\beta)$ of signals. 

The expected modulation amplitudes  become a function of the WIMP mass $A_i(m_\chi)$ since the WIMP mass $m_\chi$ determines the recoil energy spectrum.  The expected rate in bin $i,j$ then becomes:
\begin{equation}
\begin{split}
R_{i,j}^{\rm ex}(\alpha,\beta)&= \int_{t_{j}-\frac{1}{2}\Delta t_{j}}^{t_{j}+\frac{1}{2} \Delta t_{j}} \biggr(\epsilon^b_{i,j}(\alpha)\cdot (B^b_it+C^{b}_{i})\\
&+ \sigma_{\chi n} \cdot \epsilon^s_{i,j}\cdot \big( C^{s}_{i}(\beta)+ A^{s}_{i}(\beta) \cos 2\pi \small{\frac{(t-\phi)}{T}} \big) \biggr) dt,
\end{split}
\label{eq:MD}
\end{equation}
where $\phi$ and $T$ were the phase and period of the modulation and $t_{j}$ and $\Delta t_{j}$ were the time-bin's center and  width, respectively. $\sigma_{\chi n}$ is  the WIMP-nucleon cross section. Both $\epsilon^b_{i,j}(\alpha)$ and $\epsilon^s_{i,j}(\alpha)$ are the relative efficiencies for background and signal, respectively, and  are shown in Fig.~\ref{fig:sys} (left).
 To account for changing background rates from long-lived isotopes such as $^{60}$Co (t$_{1/2}=$5.27~yrs) and $^{210}$Pb (t$_{1/2}=$22.3~yrs), we added a simple linear function with  $B^b_i$ for its slope and $C_i^b$ for its constant term in the $i$th energy bin. $A^s_i(\beta)$ represents an amplitude and $C^{s}_i(\beta)$ a constant for the unmodulated component of the  signal in the $i$th energy bin after all cuts at the normalization point on day 70.
To obtain the WIMP-nucleon cross section the data were fitted in the energy range from 1.0 to 20~keV$_{\rm{ee}}$.  We assume a standard spherical isothermal galactic halo model with a most probable speed of $v_{0}$=220~km/s, Earth's 
velocity relative to the dark matter distribution of $v_{E} = 232 + 15 ~{\rm sin} 2\pi(t -\phi)/T$~km/s,  a galactic escape velocity of $v_{esc}$ = 544 km/s \cite{RAVE} and a local dark matter density of 0.3 GeV/cm$^{3}$, following \cite{Lewin}.  $T$ and $\phi$ were fixed to 365.24 and 152.5 days, respectively.
In this analysis, the signal efficiencies for each WIMP mass were estimated from MC simulations of uniformly distributed nuclear recoil events in the LXe volume.

\begin{figure}[b]
\centering
\includegraphics[width=0.5\textwidth]{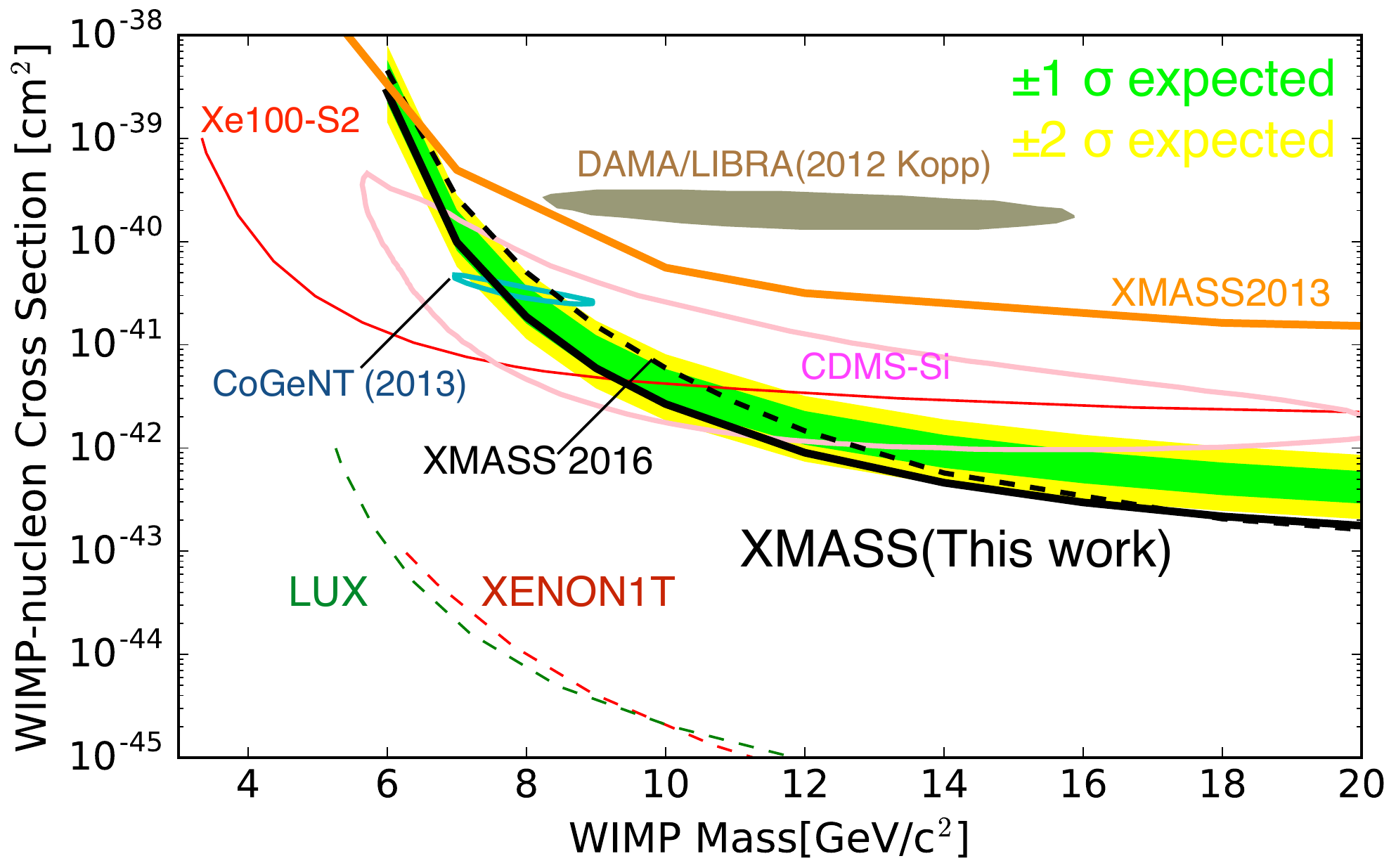}
\caption{Limits on the spin-independent elastic WIMP-nucleon cross section as a function of WIMP mass. The black solid line shows the XMASS 90\% C.L. exclusion from the  annual modulation analysis. The $\pm1\rm{\sigma}$ and $\pm2\rm{\sigma}$ bands represent the expected  90\% exclusion distributions. Limits, as well as allowed regions from other searches based on event counting, are also shown \cite{LUX, XENON1T, KOPP_DAMA, CoGeNT, CDMS-Si,XMASS1, Xe100}. }
\label{fig:MD}
\end{figure}
The best fit for an 8 GeV/c$^2$ WIMP mass had $\chi^2$/ndf =  2357/2314 and $\sigma_{{\chi}n} = (-0.7^{+ 1.0}_{- 1.7}) \times 10^{-41}$ cm$^2$.
As we found no significant signal, a 90\% C.L. upper limit on the WIMP-nucleon cross section was set for each WIMP mass.
We use the probability function $P$ defined as: 
\begin{equation}
P={\rm exp}\left(-\frac{\chi^2(\sigma_{{\chi}n})-\chi^2_{\rm min} }{2}\right),
\label{eq:prob}
\end{equation}
where $\chi^2(\sigma_{{\chi}n})$ is evaluated as a function of the WIMP-nucleon cross section $\sigma_{{\chi}n}$, while $\chi^2_{\rm min}$ is the minimum $\chi^2$ of the fit.
To obtain our 90\% C.L. exclusion upper limit $\sigma_{up}$, we used a Bayesian approach:

\begin{equation}
 \int_{0}^{\sigma_{up}} Pd\sigma_{{\chi}n}/\int^{\infty}_0 Pd\sigma_{{\chi}n} = 0.9,
\label{eq:integ}
\end{equation}
 and an upper limit of 1.9$\times10^{-41} \rm{cm}^{2} $ was derived for a WIMP mass of  8 GeV/c$^{2}$. Figure \ref{fig:MD} shows  our exclusion curve on the spin-independent elastic WIMP-nucleon cross section as a function of the WIMP mass in comparison to other experiments.
To evaluate our sensitivity for the WIMP-nucleon cross section, we carried out the  statistical test of applying the same analysis to 10,000 dummy samples with the same statistical and systematic errors as data but without any modulation following the  procedure in \cite{XMASS_MOD}.
The procedure starts by extracting an energy spectrum from the observed data. Then a toy MC simulation was carried out to produce time variations of background event rates for  each energy bin assuming the same live time as data and including systematic uncertainties.
 The $\pm1 \sigma$ and  $ \pm2\sigma$ bands in Fig.~\ref{fig:MD} outline the expected 90\% C.L. upper limit band for the no-modulation hypothesis using the dummy samples.
The result excludes the 3$\sigma$ DAMA/LIBRA allowed region as interpreted in \cite{KOPP_DAMA}.  
\begin{figure}[tp]
\centering
\includegraphics[width=0.5\textwidth]{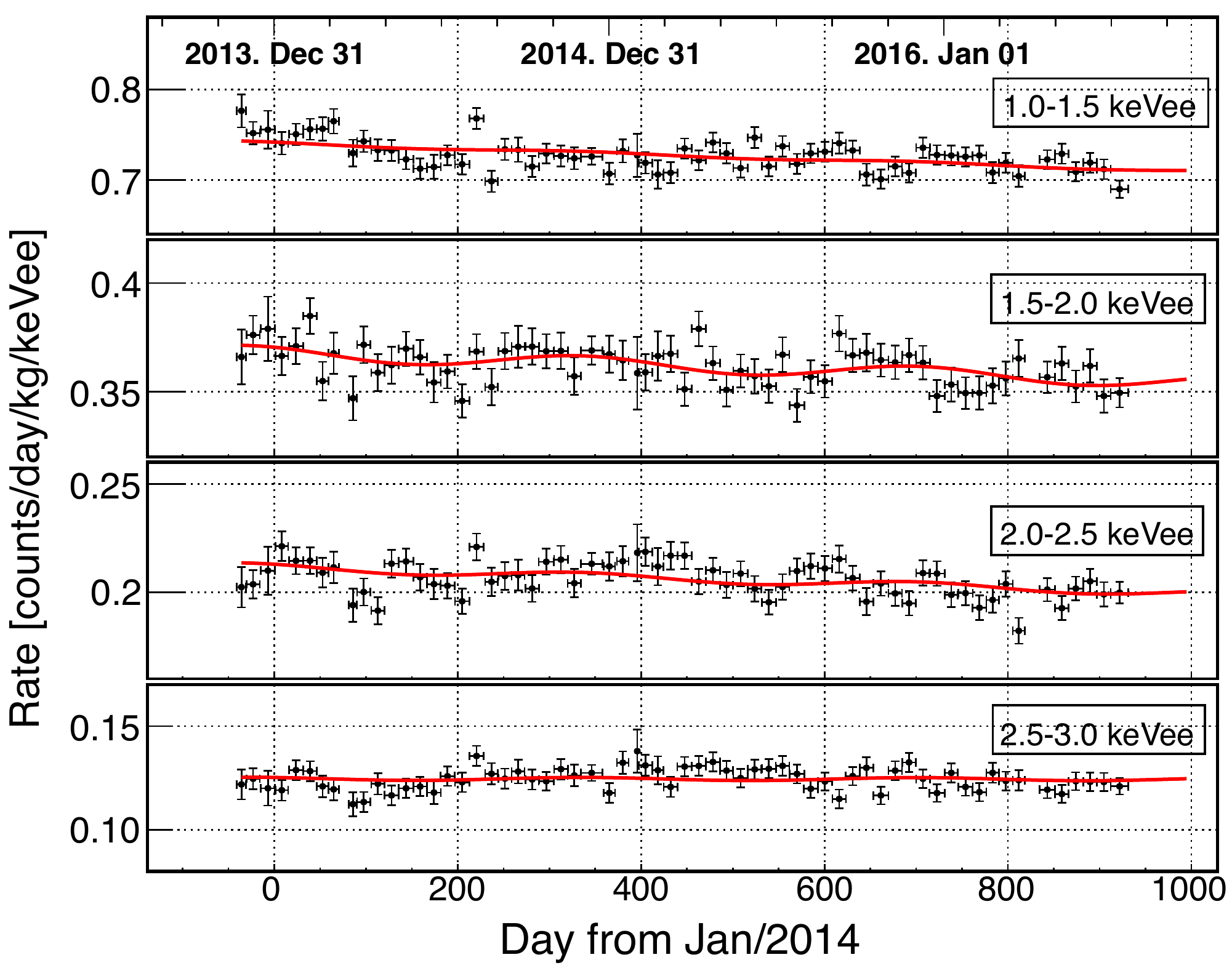}
\caption{Observed count rate as a function of time in the 1.0 -- 3.0 keV$_{\rm ee}$  energy range after correcting relative efficiency  (see text).  The black error bars show the statistical uncertainty of the count rate. The solid curves represent  the best fit result for a model-independent analysis before correcting for total efficiency.}
\label{fig:rate}
\end{figure}

\subsection{Model-independent analysis}
 For the model-independent analysis, the $\chi^2$ is expressed as
   \begin{equation}
\chi^2 = \sum\limits_{i}\limits^{E_{bins}} \sum\limits_{j}\limits^{t_{bins}} 
\left(\frac{(R^{{\rm data}}_{i,j}-R^{\rm ex}_{i,j})^2}{\sigma({\rm stat})^2_{i,j}+\sigma({\rm sys})^2_{i,j}}\right)+\alpha^{2}, 
\end{equation}

with the expected event rate written as
\begin{multline}
R_{i,j}^{\rm ex} = \int_{t_{j}-\frac{1}{2}\Delta t_{j}}^{t_{j}+\frac{1}{2} \Delta t_{j}} 
\left( \epsilon^s_{i,j} A^s_{i} \cos 2\pi \small{\frac{(t-\phi)}{T}} \right.\\
\left. + \epsilon^b_{i,j}(\alpha) (B^{b}_it + C^{b}_{i})\right) dt,
\label{eq:MI}
\end{multline}
where the free parameters $C_{i}^b$ and $A_{i}^s$ were the unmodulated event rate and the modulation amplitude without absolute efficiency correction, respectively.  In the fitting procedure,  the 1.0$-$20~keV$_{\rm ee}$ energy range was used and the modulation period $T$ was fixed to one year (= 365.24~days) and the phase $\phi$ to 152.5 days ($\sim$2nd of June)  when the Earth's velocity relative to the dark matter distribution is expected to be maximal. The observed count rate after cuts as a function of time in the energy region between 1.0 and 3.0~keV$_{\rm{ee}}$ is shown in Fig.~\ref{fig:rate}. For an easy visualization, the data points were corrected for relative efficiency based on the best-fit parameters instead of  the fitting function, therefore, the fitted line in Fig.~\ref{fig:rate} is simply a cosine plus a one-dimensional polynomial function. 

\begin{figure}[tb]
\centering
\includegraphics[width=0.48\textwidth]{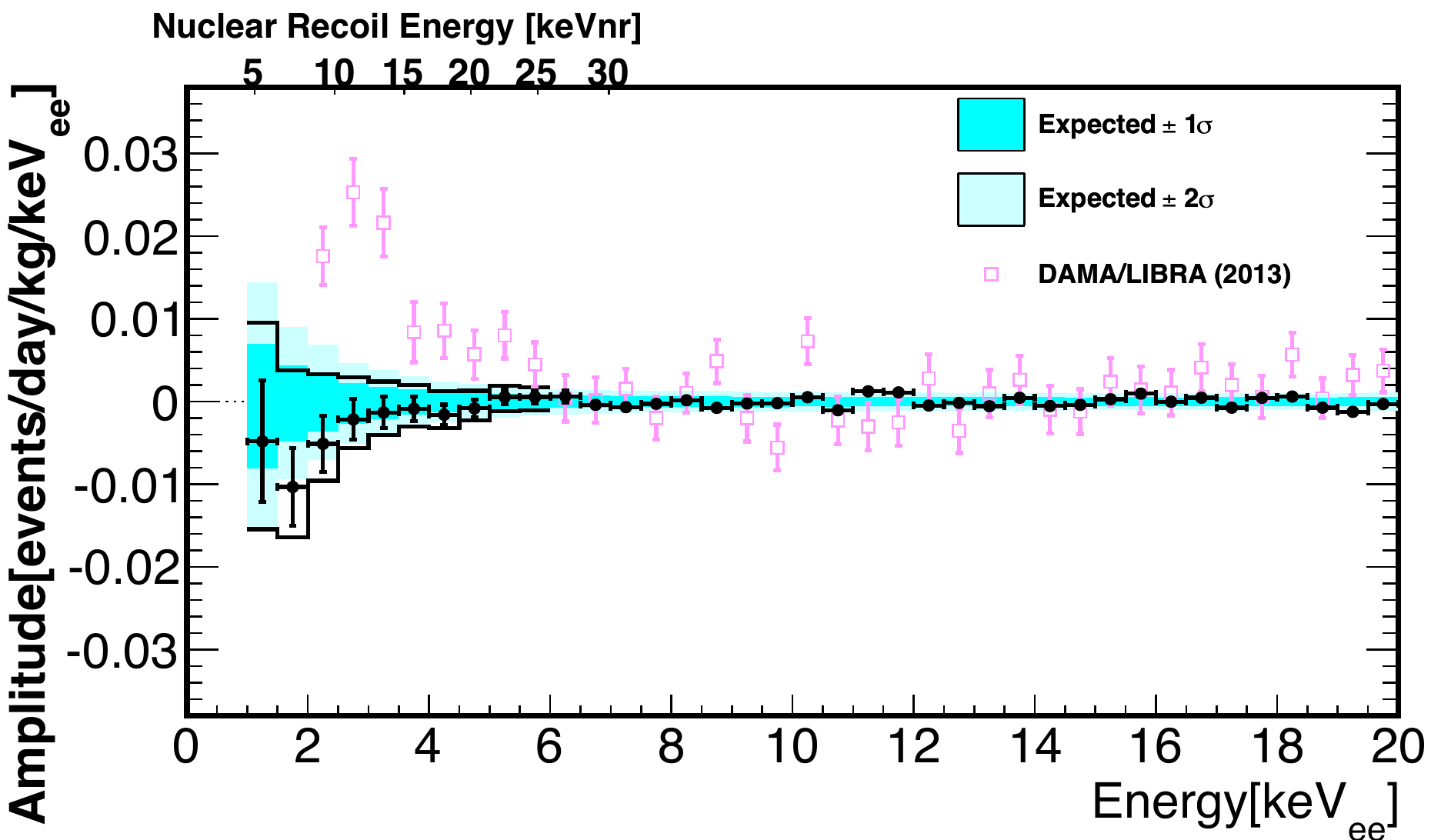}
\caption{Modulation amplitude  as a function of energy for the model-independent analyses using  the pull method (solid circle). Solid lines represent the 90\% positive (negative) upper limits on the amplitude.  The $\pm1\rm{\sigma}$ and $\pm2\rm{\sigma}$ bands represent the expected amplitude region (see detail in the text). The signal efficiency was corrected by using the curve in Fig.\ref{fig:hist} (bottom).} The DAMA/LIBRA result (square) is also shown \cite{dama}. 
\label{fig:MI}
\end{figure}

 We obtained the best-fit parameters in the energy between 1 and 20 ~keV$_{\rm{ee}}$ for the modulation hypothesis with  $\chi^2_1$/ndf = 2308/2279 and $\alpha=-0.47\pm0.15$. The result  for a null  hypothesis by fixing  $A^s_i = 0 $ was $\chi^2_0$/ndf  = 2357/2317 and $\alpha=-0.61\pm0.34$.
Figure~\ref{fig:MI} shows the best-fit amplitudes as a function of energy after correcting for efficiency by using the curve in Fig.\ref{fig:hist} (bottom).  The $\pm1\rm{\sigma}$ and $\pm2\rm{\sigma}$ bands in Fig.~\ref{fig:MI} represent expected amplitude coverage derived through the same dummy sample procedure as above.
 The hypothesis test was also done  with  these dummy samples  using their $\chi^2$ difference, $\chi^2_0-\chi^2_1$, to evaluate a $p$-value.
  This test gave the $p$-value of 0.11 (1.6$\sigma$). 
  
 As a cross-check, we also carried  this test  out for run 1 and run 2 independently in the energy region between 1 to 6 keV$_{\rm ee}$ which is almost the same as in our previous work \cite{XMASS_MOD}. Run 1 gave a  slightly higher $p$-value of 0.043 (2$\sigma$) in this analysis than in our previous one, which had a  $p$-value of 0.014. Run2 showed less than 1$\sigma$ significance.
The size of the systematic error on the amplitude was reduced from 56\% in run 1 to 22\% of total error in run 2 for the 1-1.5 keV$_{\rm ee}$ energy bin due to the stability of the PE yield as shown in Fig.\ref{fig:stability}.
 To be able to test any model of dark matter, we evaluated the constraints on the positive and negative amplitude separately in Fig.~\ref{fig:MI}.
 The upper limits on the amplitudes in each energy bin were calculated by considering only regions of positive or negative amplitude. They were calculated by  integrating Gaussian distributions based on the mean and sigma of data (=$G(a)$) from zero.  The positive or  negative upper limits are satisfied with 0.9 for  $\int_0^{a_{up}} G(a)da/\int_0^\infty G(a)da$ or $\int_{a_{up}}^{0} G(a)da/\int_{-\infty}^0 G(a)da$, where $a$ and $a_{up}$  are the amplitude and its 90\% CL upper limit, respectively.
 This method obtained a positive (negative) upper limit of $0.96~(-1.5)\times10^{-2}$ events/day/kg/keV$_{\rm{ee}}$  between 1.0 and 1.5 keV$_{\rm{ee}}$ and the limits become stricter at  higher energy.  The energy resolution ($\sigma/$E) at 1.0 (5.0) keV$_{\rm{ee}}$ is estimated to be 36\% (19\%) comparing our gamma ray calibration to its MC simulation.
As a guideline, we make the direct comparisons with other experiments not by considering a specific dark matter model. A modulation amplitude of $\sim2 \times10^{-2}$ events/day/kg/keV$_{\rm ee}$ between 2.0 and 3.5 keV$_{\rm ee}$ was obtained by DAMA/LIBRA \cite{dama}, and  XENON100 reported $(1.67\pm0.73)\times10^{-3}$ events/day/kg/keV$_{\rm ee}$ (2.0$-$5.8  keV$_{\rm ee}$) \cite{XENON_MOD}. This result corresponds to a 90\% CL upper limit (one-sided) of $2.9\times10^{-3}$ events/day/kg/keV$_{\rm ee}$.  Our study obtained a 90\%~CL positive upper limits of $(1.3-3.2)\times10^{-3}$ events/day/kg/keV$_{\rm ee}$ in the same energy region  and  gives the more stringent constraint above 3.0 keV$_{\rm ee}$  as shown in Fig.\ref{fig:MI}. This fact is important when we test dark matter models.

 \subsection{Frequency analysis}
  \begin{figure}[tb]
	\centering
	\includegraphics[width=0.48\textwidth]{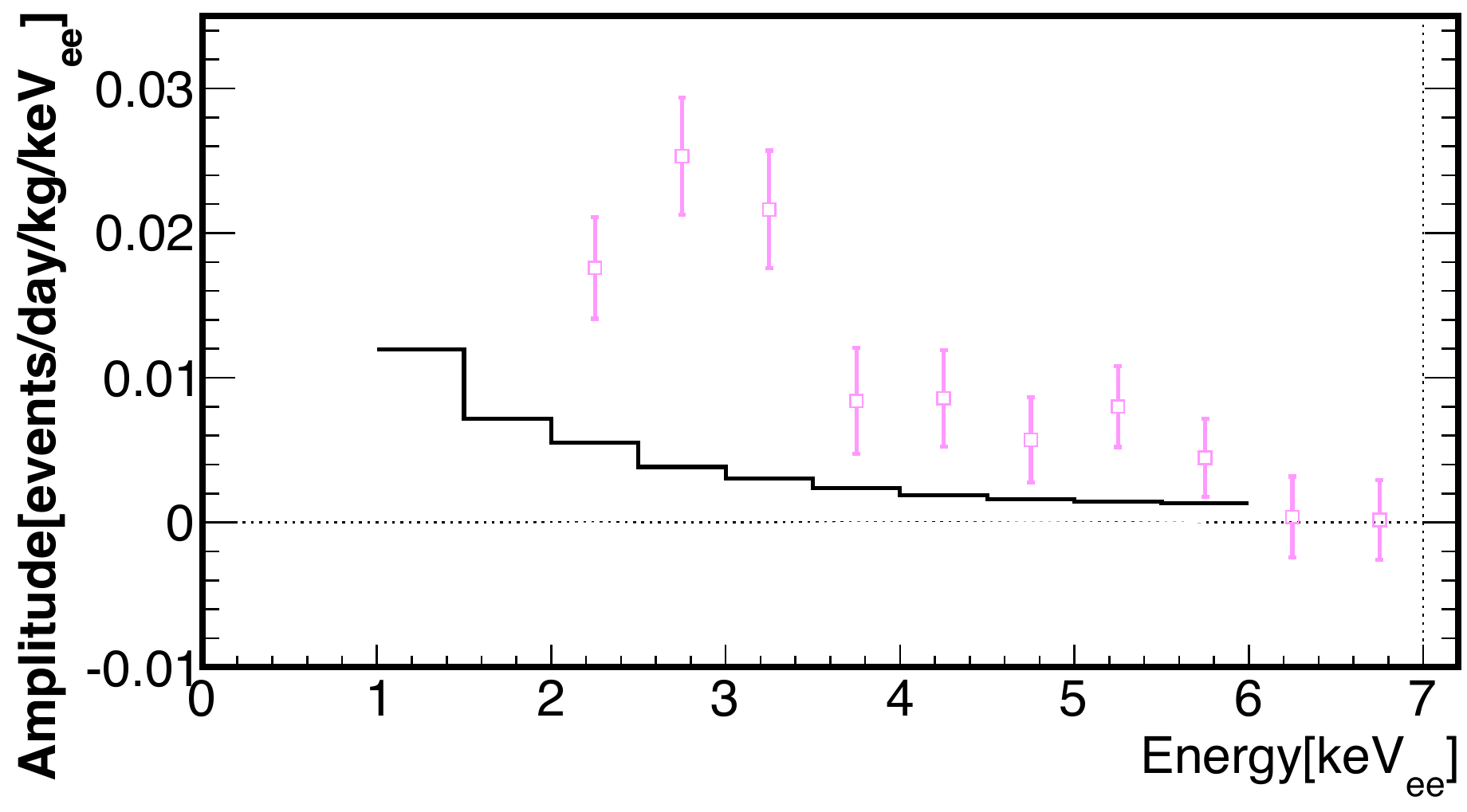}
	\caption{Input signal to test the frequency analysis together with the DAMA/LIBRA result for reference. This signal was added
		to a background dummy sample.}
	\label{fig:input}
\end{figure}
 \begin{figure}[t]
	\centering
	\includegraphics[width=0.48\textwidth]{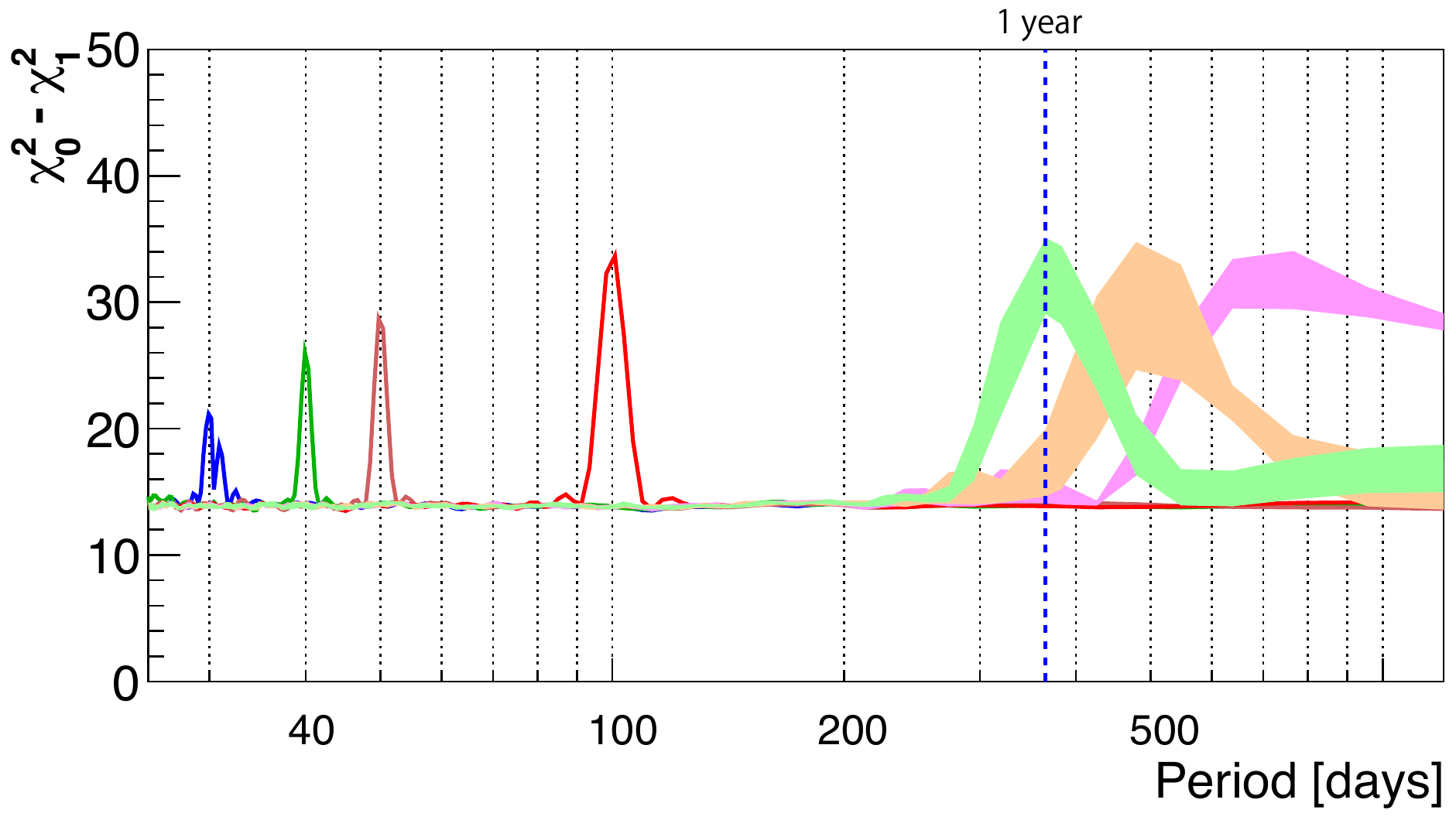}
	\caption{The mean of ($\chi^2_0-\chi^2_1$) applied to dummy samples with an artificial periodicity of $T$= 30, 40, 50, 100, 365.34, 500, and 700 days. The width of the bands reflects the time-dependent systematic error.}
	\label{fig:period}
\end{figure}
\begin{figure}[tb]
	\centering
	\includegraphics[width=0.48\textwidth]{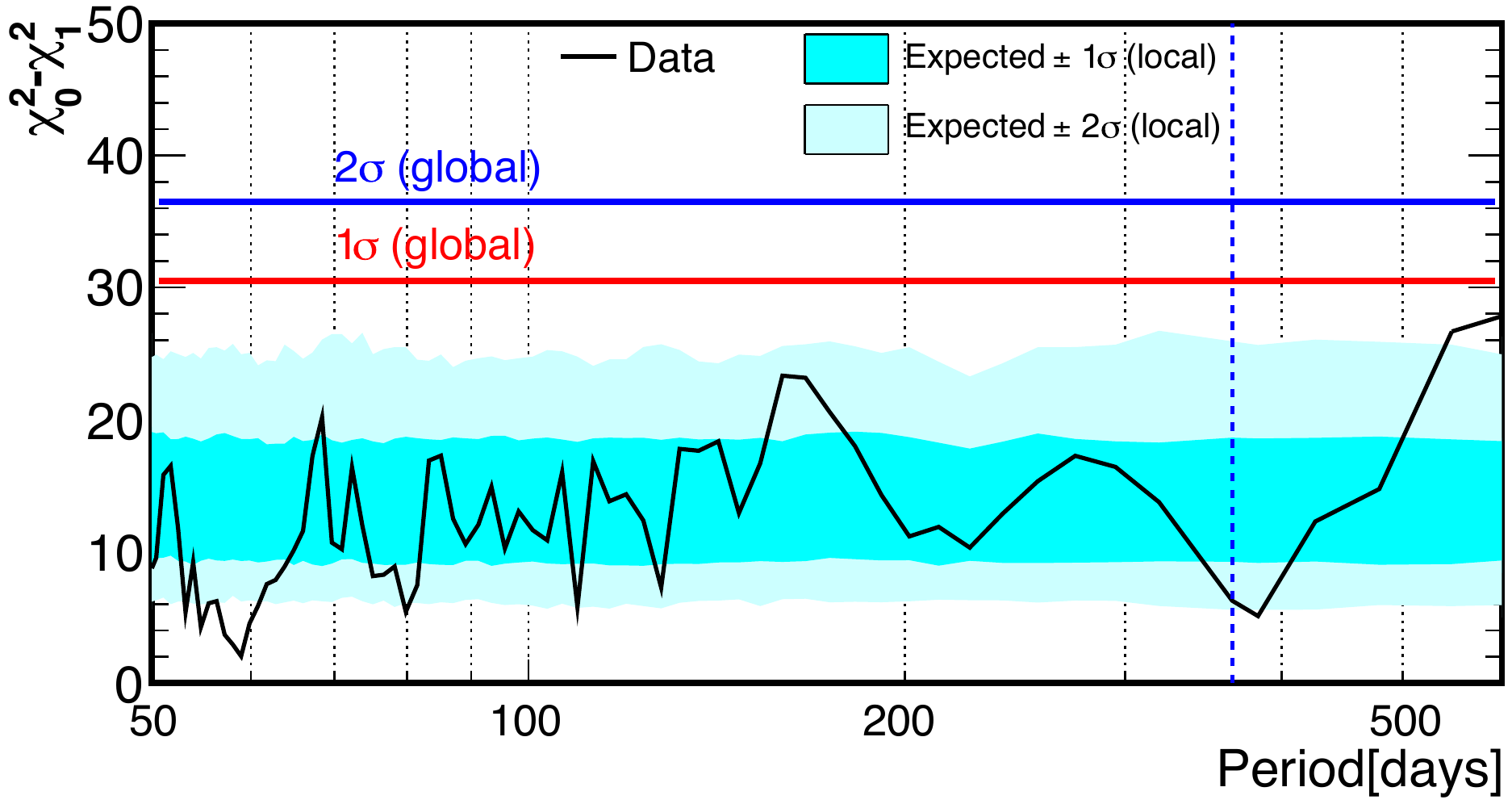}
	\caption{($\chi^2_0-\chi^2_1$)  between 50 and 600 day periods  for the 1--6 keV$_{\rm ee}$  energy range together with expected bands for $\pm1\sigma$ and $\pm2\sigma$ (local) and lines for global $1\sigma$ and $2\sigma$ significance.} 
	\label{fig:power}
\end{figure}

 To find any periodicity in the data at a low energy where a significant amplitude was observed by DAMA/LIBRA, a frequency analysis was also performed and studied in the energy range between 1$-$6 keV$_{\rm ee}$ as a function of a period.  We treated the phase $\phi$ as a free parameter, as the negative amplitude may indicate phase difference from the standard halo model. The signal strength in periodicity was calculated by the $\chi^2$ difference ($\chi^2_0-\chi^2_1$) with 11 degrees  of freedom (10 parameters come from the amplitudes in the ten 0.5 keV$_{\rm{ee}}$  bins between 1 and 6~keV$_{\rm{ee}}$ and one from the phase) as a test statistics. To demonstrate our ability to find modulation with a particular period, we use dummy samples with a simulated input signal, following \cite{XENON_MOD}.  The amplitude as a function of energy was the same as the expected amplitude distribution band in Fig.~\ref{fig:MI} and the actual amplitude in each energy bin is shown in Fig.~\ref{fig:input}. 
  The amplitude of this simulated input signal with different periodicities was chosen to reproduce the ($\chi^2_0-\chi^2_1$) of about 30 ($\sim3\sigma$)  at $T$ = 365.24 days.
   With these dummy samples for the periods of $T = $ 30, 40, 50, 100, 365.24, 500, and 700 days we tested our sensitivity. The mean of ($\chi^2_0-\chi^2_1$) is shown in Fig. \ref{fig:period}.  $T = 50$ days and longer periods show a $\chi^2_0-\chi^2_1$  of about 30 ($\sim3\sigma$), however, shorter periods lose significance as  the time-bin width was about 15 days in this analysis.
 As can be seen on the longer period side of the 700 days sample in Fig.\ref{fig:period}, periods approaching the duration of the data taking  become more difficult to distinguish from one another.
 Therefore, we tested only for periods between  50 days and 600 days in the data.  We also find that the time-dependent systematic error from the relative efficiency affects the significance of $T = 365.24$ days and longer periods. The impact was estimated by fitting the dummy sample with various phases and is shown as bands in Fig.\ref{fig:period}.
Figure \ref{fig:power} shows the result from the real science data together with the expected distribution from the dummy samples without any signal for  local significances greater than $\pm$ 1$\sigma$ and $\pm$2$\sigma$.
To check also for the `look elsewhere effect', we give the global significance (one-sided) by evaluating the maximum  ($\chi^2_0-\chi^2_1$)  in the calculated range for each sample test.  No significant periodicity was found between 50 and 600 days.

\section{Conclusions}
 In conclusion, XMASS-I with its large exposure and low energy threshold conducted an annual modulation search with 2.7 years data. For the WIMP analysis,  a 90\% CL exclusion upper limit of 1.9$\times10^{-41} \rm{cm}^{2} $ at 8 GeV/c$^{2}$ was obtained and this result excludes the DAMA/LIBRA allowed region at the 3$\sigma$ level.
 As for the model-independent case,  this analysis started from an energy threshold of 1.0 keV$_{\rm ee}$,  which is lower than that of DAMA/LIBRA and XENON100.   We did not find any modulation signal, therefore, we gave a positive (negative) upper limit for the amplitude of $0.96~(-1.5) \times10^{-2}$ events/day/kg/keV$_{\rm{ee}}$  between 1.0 and 1.5 ~keV$_{\rm{ee}}$ and $(1.3-3.2)\times$10$^{-3}$ events/day/kg/keV$_{\rm ee}$ between 2 and 6~keV$_{\rm ee}$. The significance of the modulation hypothesis was smaller than in our previous work \cite{XMASS_MOD}. As this analysis is not restricted to nuclear recoils, a simple electron or gamma ray interpretation of  the DAMA/LIBRA signal would also fall under this limit. We also did not find any particular periodicity in the data with periods between 50--600~days in the 1$-$6 ~keV$_{\rm{ee}}$ energy region.

\section*{ACKNOWLEDGMENTS}
We gratefully acknowledge the cooperation of Kamioka Mining
and Smelting Company. 
This work was supported by the Japanese Ministry of Education,
Culture, Sports, Science and Technology, 
the joint research program of the Institute for Cosmic Ray Research (ICRR), the University of Tokyo,
Grant-in-Aid
for Scientific Research, JSPS KAKENHI Grant No.~19GS0204, No.~26104004,
 and partially by the National Research Foundation of Korea Grant funded
by the Korean Government (NRF-2011-220-C00006).





%

\end{document}